\documentclass{pasj00}
\draft

\usepackage{multirow}
\usepackage{url}

\begin{document}
\SetRunningHead{K.Arimatsu et al.}{the AKARI/FIS all-sky maps for stacking analysis}

\title{Point source calibration of the {\it AKARI}/FIS all-sky survey maps for staking analysis}

\author{
  Ko       \textsc{Arimatsu}\altaffilmark{1,2}
  Yasuo    \textsc{Doi}\altaffilmark{3}
  Takehiko \textsc{Wada}\altaffilmark{1}
  Satoshi  \textsc{Takita}\altaffilmark{1}
  Mitsunobu \textsc{Kawada}\altaffilmark{1}
  Shuji    \textsc{Matsuura}\altaffilmark{1}
  Takafumi \textsc{Ootsubo}\altaffilmark{4}  
\& Hirokazu \textsc{Kataza}\altaffilmark{1}
}
\altaffiltext{1}{Institute of Space and Aeronautical Science,
Japan Aerospace Exploration Agency, 3-1-1 Yoshinodai, Chuo-ku, Sagamihara, 
Kanagawa 229-8510, Japan.}
\email{arimatsu@ir.isas.laxa.jp}
\altaffiltext{2}{Department of Astronomy, Graduate School of Science, The University of Tokyo, 7-3-1 Hongo, Bunkyo-ku, Tokyo 113-0033, Japan}
\altaffiltext{3}{Department of Earth Science and Astronomy, the University of Tokyo, Tokyo 153-8902, Japan.}
\altaffiltext{4}{
Astronomical Institute, Tohoku University, Aramaki, Aoba-ku, Sendai 980-8578, Japan}

%


\maketitle

\begin{abstract}
Investigations of the point spread functions (PSFs) and flux 
calibrations for stacking analysis have been performed with 
the far-infrared (wavelengths range of 60 to 140\,$\mu$m) all-sky maps taken by 
the Far-Infrared Surveyor (FIS) onboard the AKARI satellite.
The PSFs are investigated 
by stacking the maps at the positions of standard stars with their fluxes 
of $0.02-10$ Jy.
The derived full widths at the half maximum (FWHMs) of the PSFs 
are $\sim 60\arcsec$ at 65 and 90\,$\mu$m 
and $\sim 90\arcsec$ at 140\,$\mu$m, which are much smaller than that 
of the previous all-sky maps obtained with IRAS ($\sim 6\arcmin$).
Any flux dependence in the PSFs is not seen on the investigated flux range.
By performing the flux calibrations, we found that absolute photometry for faint sources 
can be carried out with constant calibration factors, which range from 0.6 to 0.8. 
After applying the calibration factors,
the photometric accuracies for the stacked sources
in the 65, 90, and 140\,$\mu$m bands 
are 9, 3, and 21\,\%, respectively, 
even below the detection limits of the survey.   
Any systematic dependence between the observed flux and model flux 
is not found.
These results indicate that 
the FIS map is a useful dataset 
for the stacking analyses of faint sources at far-infrared wavelengths.

\end{abstract}

\newpage

\section{Introduction}

\label{intro}

{\it AKARI} is the first Japanese satellite that carried out 
mid- to far-infrared (hereafter MIR and FIR, respectively) all-sky survey
\citep{murakami07}.
The {\it AKARI} all-sky survey in FIR wavelengths were performed 
from May 8 2006 to August 28 2007
with the Far-Infrared Surveyor (FIS, \cite{kawada07})
in four bands, covering 50-180\,$\mu$m wavelength range. 
The FIS scanned more than 98\,\% of the sky, and 
provides the FIR all-sky dataset with the highest spatial resolutions 
and sensitivities \citep{kawada07}.

From the FIS all-sky data, the FIR bright source catalogue (BSC) is released \citep{yamamura10}. 
In addition, the diffuse maps will be publicly available from the data archive server. 
The diffuse map will enable us to perform studies of 
an arbitrary set of objects.
However, individual objects are usually faint at FIR wavelengths, 
and severely contaminated with diffuse emission 
from interstellar dust and background unresolved sources.

Stacking analysis is a useful technique
to investigate FIR 
spectral trends and spatial structures of these faint sources.
The confusion noise can be reduced by stacking the map at the positions of 
the sample objects. 
With the staked image, 
we can perform investigation of the FIR photometry and spatial structures
 of the sample with higher sensitivities than that performed with individual images.
Therefore the stacking technique has been extensively adopted   
for the statistical FIR studies for faint sources, such as known
 normal galaxies and quasars \citep{kashiwagi12}, 
 galaxy clusters \citep{giard08}, 
and mid-infrared extragalactic sources in deep field survey areas \citep{jauzac11}. 
These studies have succeeded in investigating FIR spectral and spatial features
of the faint sources with fluxes even below the detection limits.

This paper focuses on the investigations of point spread functions (PSFs) and flux calibrations
for sources with fluxes comparable or below the detection limits, 
which are major objectives for the stacking analyses.
In the stacking analyses, 
photometric accuracy and the PSFs of the images
are assumed to be well calibrated, and 
have no systematic change throughout the considering flux range.
However, the characteristics of point sources in the far-infrared data is generally very complex,
mainly due to slow transient response of detectors \citep{shirahata09}, and 
diffraction and scattering inside the detector arrays and optical instruments \citep{arimatsu11}.
These instrumental effects result in the differences between point source and diffuse calibration.
The FIS all-sky map is calibrated in the earlier stage of the data processing 
by diffuse sources, such as zodiacal and interstellar dust emission \citep{matsuura11}.
Additional investigation and calibration of the maps are required for point source studies.
Furthermore, the photometric performance and the PSFs can 
systematically change in the FIS maps against the source fluxes, 
because the detector response is flux dependent \citep{shirahata09}. 
Therefore, we need to check the flux-dependent characteristics of the FIS maps. 

In the present study, we investigate the {\it AKARI} FIS diffuse maps 
at the N60 (65\,$\mu$m), Wide-S (90\,$\mu$m), and
Wide-L (140\,$\mu$m) bands. 
The $5 \sigma$ detection limits for individual point sources in the maps are
2.4, 0.55, and 1.4\,Jy at 65, 90, and 140\,$\mu$m, respetively \citep{kawada07}. 
The N160 (160\,$\mu$m) band map is not investigated 
because the sensitivity is too low 
($5 \sigma$ detection limit: 6.3 Jy, \cite{kawada07}) for the faint source studies.

As calibration sources, we adopted infrared standard stars proposed by \citet{cohen99}
with a flux range from 0.05 to 10 Jy.
To investigate spatial and photometric properties of the faint stars, 
we stacked the FIS maps at their positions. 
By using the standard stars and the staking method,
we present  the characteristics of the point sources in the FIS maps 
with fluxes down to ten times lower than
 the detection limits for individual objects.
These investigations provide not only useful information for users of the {\it AKARI}/FIS maps,
but also demonstrate a investigation and calibration scheme 
for the general FIR stacking analyses.

In this paper, 
section 2 describes a selection of standard stars 
and the stacking method used for the present study.
In section 3, we report the results of the stacking analysis, 
trying to constrain the properties of point spread functions and
calibration factors for point sources with different fluxes.
Our results are discussed in section 4, and
section 5 offers a short summary.

\section{Data analysis}

\subsection{FIS all-sky diffuse map}

In the preset study, we have used the FIS all-sky diffuse maps version 130401, 
which went through data reduction procedures that apply to the publicly released dataset. 
The all-sky maps are produced from time-series scan data obtained during the all-sky observations.
In the observations, the detector sweeps the sky in a direction almost perpendicular to the ecliptic latitude
with a scan speed of $3\arcmin.6\, {\rm sec^{-1}}$.
During the scan, the integrated detector signals are continuously read out as the time-series data.
To construct maps from the time-series data, the following two pipelines are carried out.
First of all, the obtained data are preprocessed 
with the FIS pipeline tool originally optimized for bright point source extraction \citep{yamamura10}.
Then, an additional pipeline is used for depicting diffuse emission structures accurately.
This pipeline consists of transient response correction of the detector signal, 
subtraction of zodiacal emission,  image processing, subtracting the stripe-like patterns, 
and recursive deglitching and flat-fielding referring processed image.
Through these procedures,
the all-sky map is produced with a sample scale of $15\arcsec$ 
and units of surface brightness in ${\rm MJy\, sr^{-1}}$.
The derived map is compared with 
the COBE/DIRBE Zodi-Subtracted Mission Average data. 
The correction factor is derived from the comparison 
and adopted for the procedures recursively.
The final map is thus confirmed to be well calibrated for the diffuse emission (Takita et al. in preparation).


\subsection{Data sets}
\label{sec_selection}

As already described in section \ref{intro}, calibration stars are selected 
from the infrared standard star catalogue 
proposed by \citet{cohen99}.
This catalogue includes 422 giant stars with spectral types of K0 to M0.
In their study, these stars have been studied extensively 
using the optical and infrared observation data.
\citet{cohen99} confirmed that the catalogued stars are well isolated, and the flux of each star is more than 20 times higher
than the total flux from nearby sources located within a radius of $6\arcmin$ at the mid-infrared
wavelengths. 
The infrared model spectrum is derived for each star 
based on the photosphere emission templates scaled by photometric and spectroscopic observation results, 
which are concluded to be reliable with accuracies better than 6\,\%.
However, these stars were checked in the FIR wavelengths range 
only with the IRAS data.
These stars can have excess emission from their circumstellar 
dust at FIR
that was not able to be detected with IRAS. 
We thus reject stars with possible FIR excesses by the following criteria. 

For the calibration stars 
with their model fluxes less than 1.0 Jy at 90\,$\mu$m,
photometric variability is investigated 
using the SIMBAD database (\url{http://simbad.u-strasbg.fr/simbad/}),
and variable stars are rejected from the stacked targets. 
Several positions of the stars are severely contaminated with background interstellar dust emission,
and we exclude positions with the background surface brightnesses of 
more than $20\, {\rm M\,Jy\, sr^{-1}}$ 
at the FIS 90\,$\mu$m band from the following analysis.
Aperture photometry is performed at the individual positions 
of the stars on the 90\,$\mu$m maps with an aperture radius of $90\arcsec$.
We compare the observed fluxes with those expected from the model spectrum
by \citet{cohen99},
and five stars with fluxes $5 \sigma$ higher than the expected fluxes are rejected. 
A search of the Cohen catalogue that satisfied these criteria produces 352 stars.
Several images are severely noisy with speckles, 
which are thought to be due to instabilities of the detector response 
after cosmic ray hitting. 
Therefore we exclude 14, 6, 18 stars from standards at 65, 90, and 140\,$\mu$m, respectively.

Number of the selected stars in each expected flux bin are presented in table~\ref{taba2}.
The expected flux is derived by convolving the model spectrum of each star
with the spectral responsivity function for the FIS bands.
The expected fluxes range from 0.1 to 19, 0.06 to 12, and 0.02 to 3.7\,Jy 
in the 65, 90, and 140\,$\mu$m bands, respectively.
Note that a different flux bin is adopted at the $140\,{\rm \mu m}$ band 
to obtain the flux-dependent trend with sufficient sensitivities.


\subsection{Stacking Analysis}

We stack the all-sky maps centered at the positions of the selected stars 
over their appropriate flux bins. 
In this procedure, we evaluate the pixel value on $3\arcsec.75 \times 3\arcsec.75$ pixels
over $10\arcmin \times 10\arcmin$ images aligned in scan direction
by cloud-in-cell interpolation of the four nearest neighbors in the original pixels.
Then the median of all pixel values is estimated as the background 
surface brightness and subtracted from each image.
After that, these images are combined by averaging to make a stacked image.
In this procedure, pixel values in each image are scaled by the expected flux so that
the brightness levels of the images are aligned with each other. 
We repeat this procedure iteratively and clip regions 
in each image with absolute pixel values more than
$4 \sigma$ above the processed image when re-stacking.

\section{Results} 

\subsection{Point Spread Functions}

\label{resluts1}
\begin{figure}
   \includegraphics[scale=0.6]{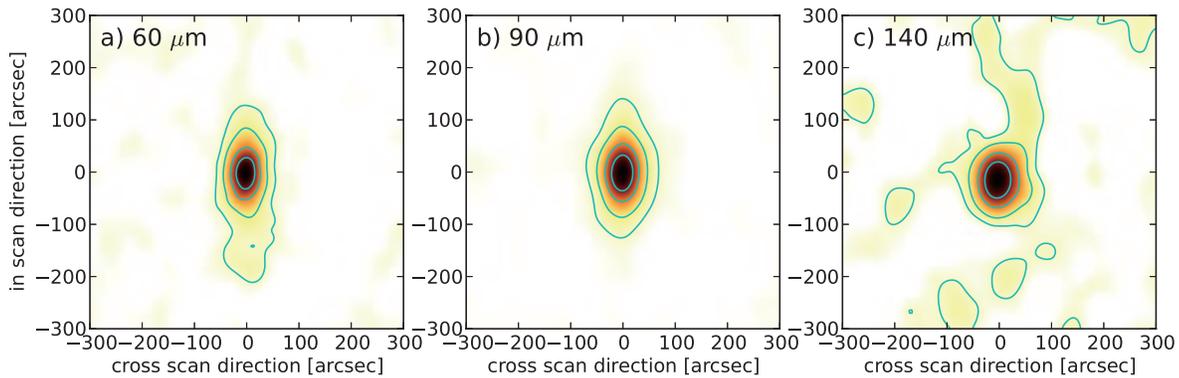}
   \caption{Obtained PSFs for the FIS $65, 90, {\rm and} \, 140\, {\rm \mu m}$ bands. 
     Contours are at 75\%, 50\%, 25\% and 10\% of the peak pixel value.}
   \label{figa2}
\end{figure}

Figure~\ref{figa2} shows the obtained PSFs derived by 
averaging 
49, 97, and 9 standard stars 
with their fluxes above the $2.5 \sigma$ detection limits,
which corresponds to 1.2, 0.28, and 0.7 Jy
at the FIS 65, 90, and 140\,$\mu$m bands, respectively.
The background emission severely contaminates the obtained image 
at the 140 \,$\mu$m band, where the fluxes of the standard stars are relatively low.
At the 65 and 90\,$\mu$m bands, PSFs are elongated in the scan direction 
(corresponds to the vertical direction in the figure).
The full widths at the half maximum (FWHMs) of the averaged radial, in-scan, 
and cross-scan directions are derived from the obtained PSFs,
which are listed in table~\ref{taba3}. 
The average FWHMs are less than $90\arcsec$ at the three bands.
Especially at the 65 and 90\,$\mu$m bands, the in-scan FWHMs are 
much larger than the cross-scan FWHMs.
A possible reason of this asymmetry is 
that the transient response of the FIS detector is 
not perfectly corrected in the data reduction procedure, 
and the latency of signal is detected as the prolonged PSF features.
In fact, the all-sky maps are constructed from the observations
performed with two different scan directions, 
from North ecliptic pole to South ecliptic pole and vice versa. 
The prolonged feature can thus be axially asymmetric for individual sources in the maps.
The stacked images are made by simply averaging 
both types of observation data by approximately the same number, 
and the asymmetry can be missed in the present study.

\begin{figure}
\includegraphics[scale=0.6]{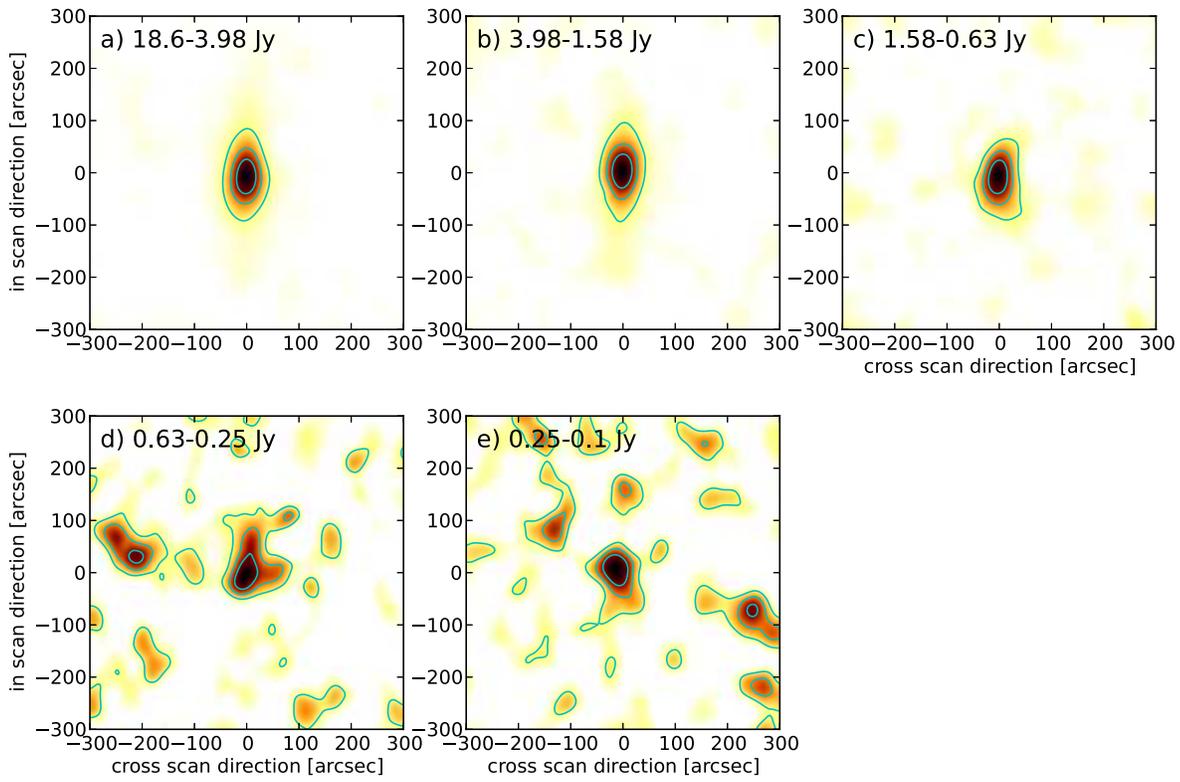}
\caption{Stacked images of standard sources for the FIS $65\, {\rm \mu m}$ band 
with different flux ranges. 
Contours are at 70\%, 50\%, and 30\% of the central pixel value.
Stacked sources in the panels d and e are severely contaminated 
with background fluctuations.
The $1\sigma$ background fluctuation levels are 3.7, 5.1, 7.6, 23, and 24\%
of the peak intensity of the central stacked source for a, b, c, d, and e, respectively.}
\label{figa3}
\end{figure}

\begin{figure}
\includegraphics[scale=0.6]{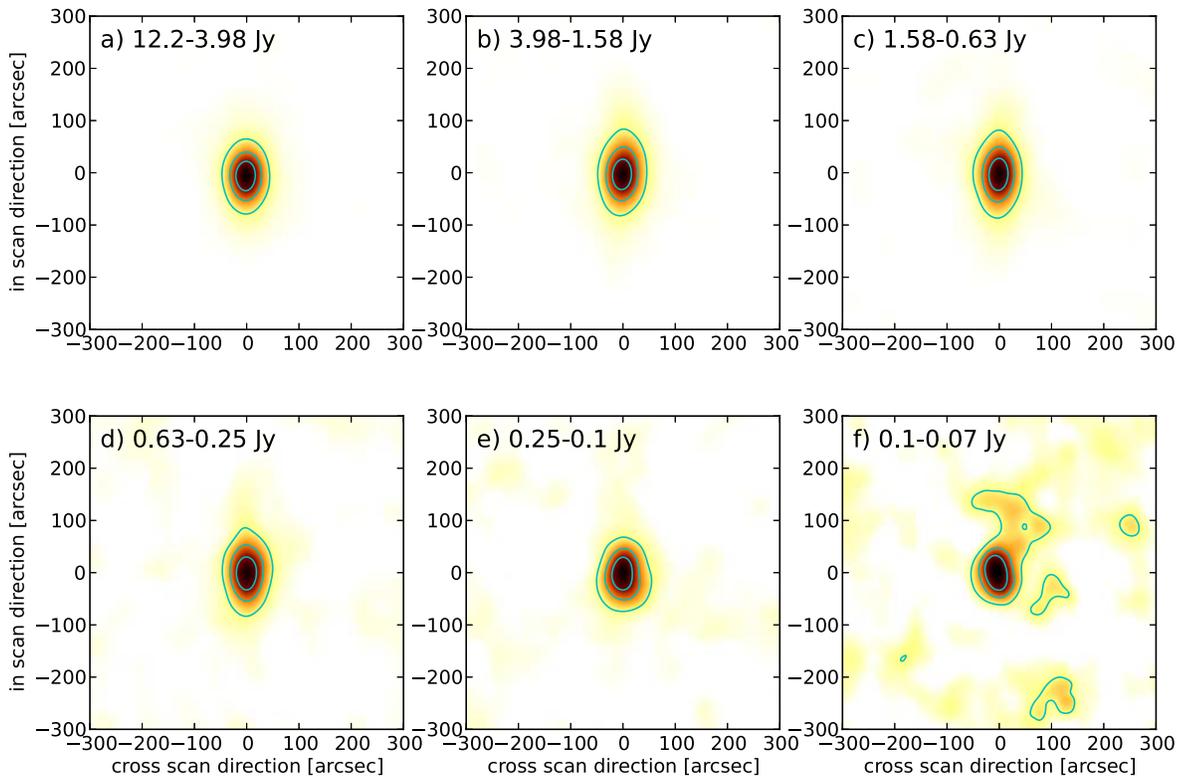}
\caption{Same as figure~\ref{figa3}, but for the $90\, {\rm \mu m}$ band.
The $1\sigma$ background fluctuation levels are 3.5, 3.8, 4.1, 6.4, 6.8, and 13\%
of the peak intensity for a, b, c, d, e, and f, respectively.}
\label{figa4}
\end{figure}

\begin{figure}
\includegraphics[scale=0.6]{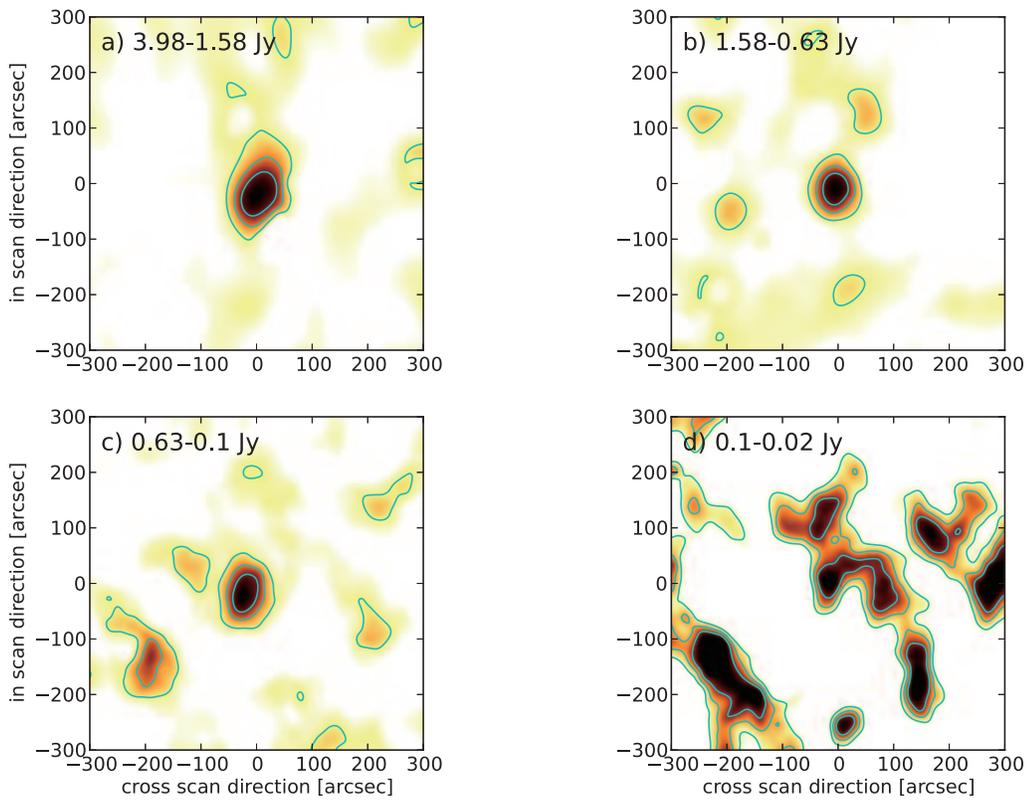}
\caption{Same as figure~\ref{figa3}, but for the $140\, {\rm \mu m}$ band.
The $1\sigma$ background fluctuation levels are 11, 14, 22, and 59\%
of the peak intensity for a, b, c, and d, respectively.
The stacked source in the panel d is severely contaminated with background structures.}
\label{figa5}
\end{figure}

 
To investigate the characteristics of the PSFs for point sources with different fluxes, 
we stack the standard sources over their appropriate flux bins. 
Panels in figures~\ref{figa3}, \ref{figa4}, and ~\ref{figa5} show 
the stacked images of standard sources 
with different flux bins at the 65, 90, and 140\,$\mu$m bands, respectively. 
All of the panels at the 90\,$\mu$m (figure~\ref{figa4}) clearly show the stacked sources 
at the image center. 
At the 90\,$\mu$m band,
the shape of the stacked sources for each flux bin follows the PSF (figure~\ref{figa2}b).
On the other hand, the intensity distributions in the images of two 
and three fainter groups at the 65 and 140\,$\mu$m bands
do not seem to follow the PSF profiles (figures~\ref{figa3}d, e and \ref{figa5}d).
This is because they are severely contaminated with background fluctuations, the fluctuation levels of which are 
larger than 20\,\% of the peak intensity of the central stacked source. 
In any case, no flux dependent trend is seen in these images.

\begin{figure}
\includegraphics[scale=0.7]{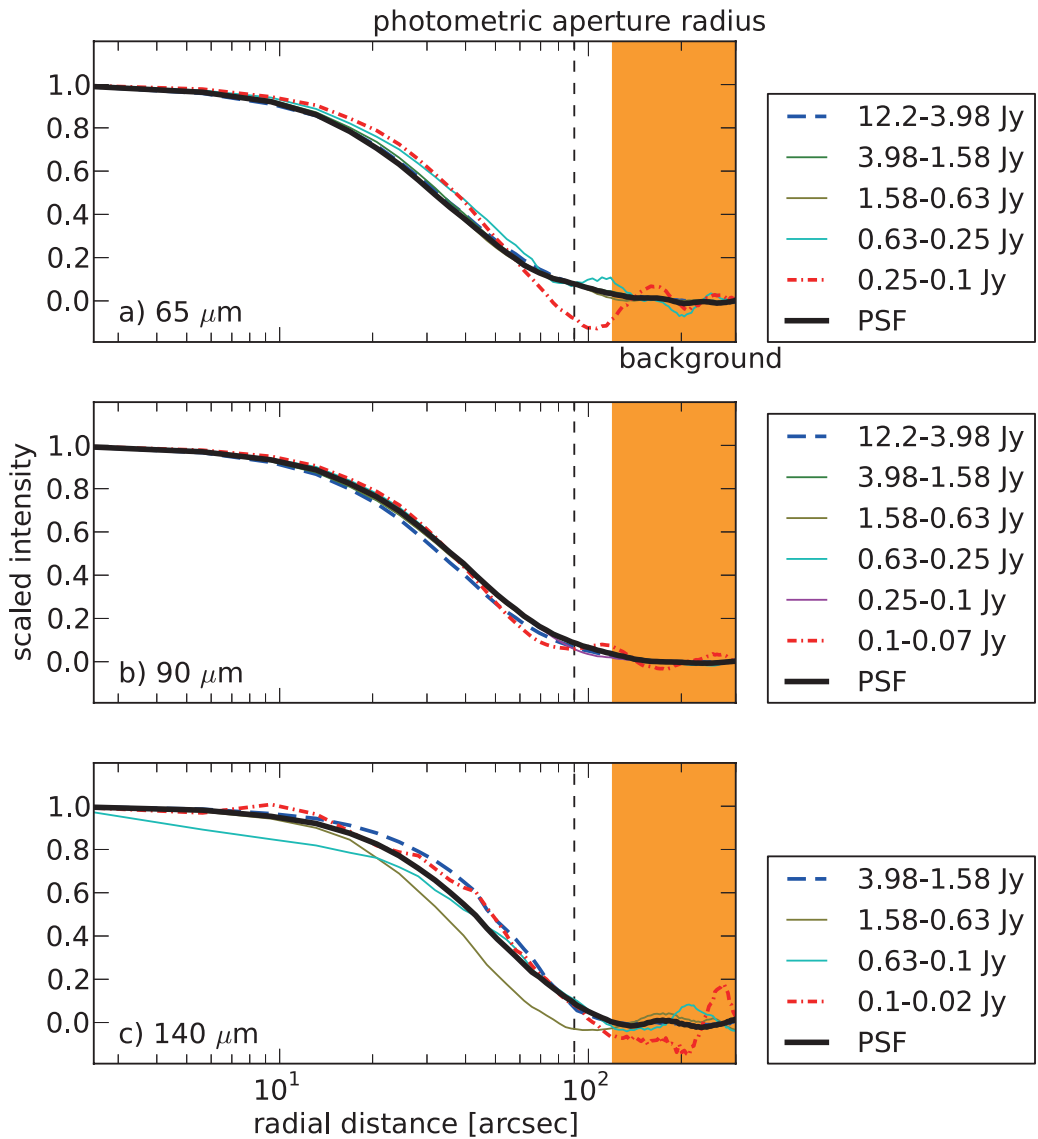}
\caption{Normalized radial intensity profiles of a point source derived by 
the stacking image at the FIS a) 65, b) 90, and c) 140\, ${\rm \mu m}$ bands.  
The lines in the individual figures are the profiles for the standard stars with different fluxes. 
The bold dashed and dot-dashed line in each panel 
are the stars for the maximum and minimum flux groups, respectively.
The bold solid lines represent the profiles of the PSFs shown in figure~\ref{figa2}.
An aperture radius ($90\arcsec$)  for photometry is shown as the vertical dashed line, 
and the background annulus ($120\arcsec -300\arcsec$) is also shown
as the orange area.}
\label{figa6}
\end{figure}

To make a further constraint on the PSF variations, 
the stacked images are azimuthally averaged to derive radial trends. 
The derived radial profiles for the bright 
(dashed line) to faint sources (dot-dashed line) are compared 
with the PSFs in figure~\ref{figa6} (thick-solid line).
These profiles are surprisingly stable over the flux range 
of the calibration stars at 90\,$\mu$m. 
Small ($\sim 10\,\%$ of the peak intensity) fluctuations can be seen 
in the profiles of the fainter groups
at the 65 and 140\,$\mu$m bands (figures~\ref{figa6}a and c) bands, but these are not systematic throughout the flux range.
These trends can be described by contamination of the background fluctuations 
seen in the stacked images (figures~\ref{figa3}d, e and ~\ref{figa5}c, e).
In conclusion, the deviations exhibit no clear systematic trend against 
the fluxes.

%
%
%

\subsection{Absolute Flux Calibration}
As already described in section \ref{intro}, 
the photometric calibration of the FIS all-sky diffuse maps are achieved with the 
diffuse sources.
In order to derive the calibration factor for point-source photometry, 
we compared the observed fluxes obtained by aperture photometry with their model prediction.
Aperture photometry of stacked point sources is performed on the stacked images 
with the aperture of $90\arcsec$ in radius, which is presented in figure~\ref{figa6}. 
The sky background is measured in the annulus with the inner and outer radii 
of  $120\arcsec$ and $300\arcsec$.
Uncertainties of the fluxes are derived from the scatter of the values 
measured in randomly selected areas with the same aperture radius of 
$90 \arcsec$ in the background annulus.

%

\begin{figure}
\includegraphics[scale=0.7]{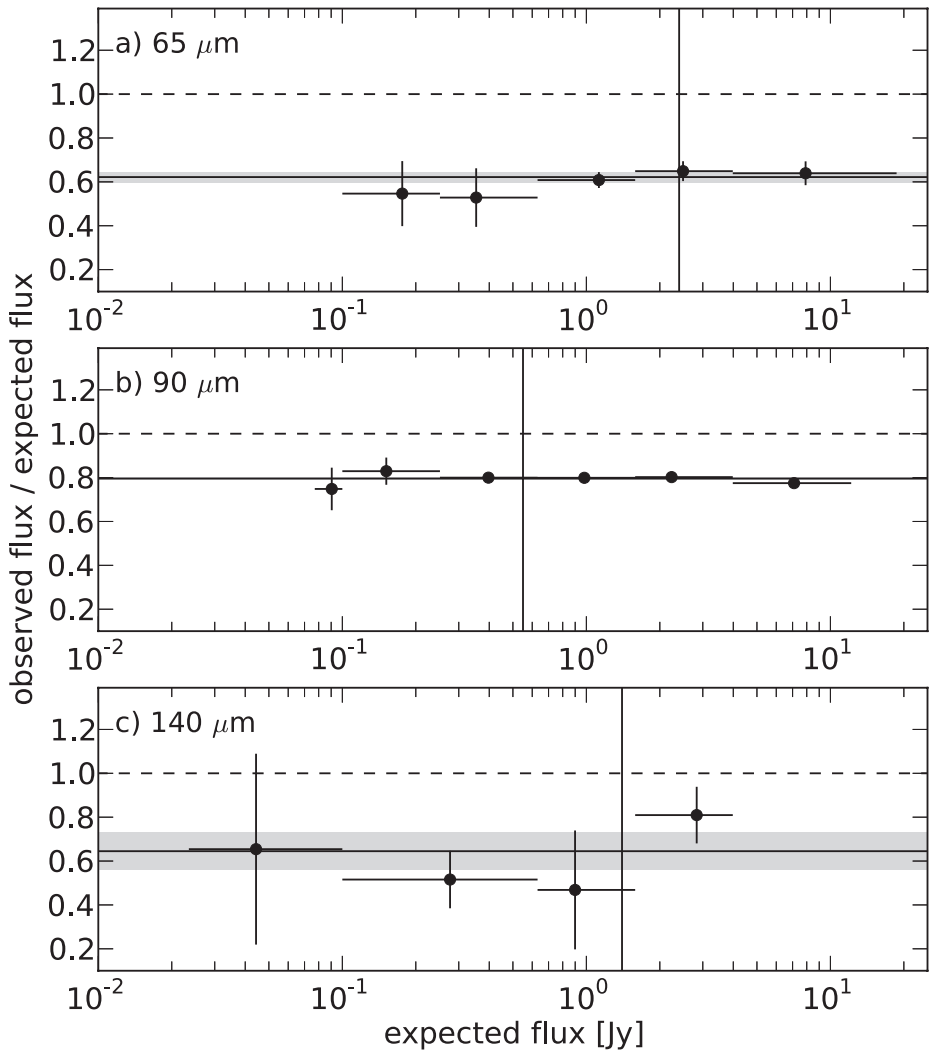}
\caption{The observed-to-expected flux ratio as a function of the expected flux at 
the FIS a) 65 , b) 90, and c) 140\, ${\rm \mu m}$.
The error bars on the $y$-axis represent the $1 \sigma$ photometry uncertainties. 
The solid lines with gray areas are the weighted averages with the $1 \sigma$ uncertainties.
The dashed lines indicate a linear relation.
The vertical line represents the $5 \sigma$ detection limit of the All-Sky Survey at each band 
\citep{kawada07}
\label{figa8}}
\end{figure}

Figure~\ref{figa8}  shows the observed-to-expected flux ratio for the stacked calibration standards
as a function of the expected flux in each flux bin.
The ratio between the observed and expected fluxes represents
the calibration factor for point-source photometry.
The observed fluxes are always lower than the expected ones at the three bands.
Possible reasons for the trend are twofold. 
One is that the observed fluxes are underestimated 
due to a extended PSF component outside the aperture radius (e.g., \cite{arimatsu11}).
This component can be very faint and difficult to detect from the PSFs with the insufficient signal-to-noise ratios
used in the present study.
Another is that the sensitivity for point sources is lower than for diffuse sources.
This trend is possible because the signal from the point sources can be 
missed due to the slow transient of the detector \citep{shirahata09}.

The flux ratios do not show any flux-dependent trend, 
and can be approximated by a constant value.
The weighted means of the ratios are derived to be 
$0.622\pm0.024$, $0.796\pm0.006$, and $0.645\pm0.085$, 
for the 65, 90, and 140\,$\mu$m bands, respectively.
The deviations of the observed fluxes from the weighted means 
are 9.0, 3.2, and 21\,\% for the 65, 90, and 140\,$\mu$m, respectively.
The 140\,$\mu$m value is relatively higher than the other two bands.
We should note that these values include photometric uncertainties
as well as the variability of the calibration factors.  
The derived values thus indicate the maximum deviations of the 
calibration factors from the constant approximation.
Taking the large error bars of photometry, as shown in figure~\ref{figa8}c,  
into consideration, the deviation at the 140\,$\mu$m is mainly
due to the large flux uncertainties. 
These results present photometry for faint sources can be performed  
with the calibration factors approximated to be constant.

%

\section{Discussions}

\subsection{Comparison with the Pointed Observations}

In addition to the all-sky survey investigated so far,
{\it AKARI}/FIS had performed slow-scan observations for the pointing mode, 
and their data characteristics were investigated in previous studies.
In the previous studies for the slow-scan observations,
the observed-to-expected flux ratios 
show a clear flux-dependence 
at the 65 and 90\,$\mu$m bands in the pointed observations \citep{shirahata09}.
On the other hand, the flux dependence of the imaging performance
is not confirmed for the FIS all-sky maps. 
The observed trend was thought to be due to the slow transient response of 
the FIS detector, which is also used for the all-sky survey.
The previous results seem to be inconsistent with the present ones.

In order to explain the discrepancy, 
we should firstly take the differences of the scan speeds for these 
observation modes into account.
During the slow-scan observations, the detector sweeps 
the sky with a
scan speed of 8 or $15\arcsec\, {\rm sec^{-1}}$, 
and the integration times of the point sources are  
9.1 or 4.6 seconds, respectively, 
assuming a angular size of the point sources is comparable to the FWHM of the 
90\,$\mu$m band PSF (73\arcsec, see table~\ref{taba3}).
These timescales are comparable to duration time  
of the slow transient response  (10-30 seconds), which are flux-dependent.
The observed flux ratio thus shows the clear flux-dependence.
On the other hand, the detector sweeps the sky with a much higher scan speed 
($3\arcmin.6\ {\rm sec^{-1}}$) in the all-sky observations.
A point source is scanned in only about 0.38 second, and the output signals 
are dominated by the initial pulse signals with duration time $\sim 0.2$ second
\citep{shirahata04}. 
According to \citet{shirahata04}, the intensities of the pulse
linearly depends on the stepwise input signal variations, 
and can purely be interpreted as the source fluxes. 
Therefore the slow transient effect can be less severe for the all-sky mode.

Another possible reason for the inconsistency comes from the 
differences of the fluxes of the objects selected for the standards. 
According to the empirical model for the FIS detector established by \citet{kaneda02}, 
the slow transient response should depend on the total photo-current, 
which includes background light and signal from the target.
Since the previous study used only the bright sources as calibration objects, such as 
asteroids and bright stars, the total fluxes are dominated by the source flux. 
Therefore the flux-dependence was clearly seen in the previous study.
In contrast, the faint sources are used in our investigations, 
and the total fluxes are dominated by the background light.
It is thus natural the flux ratio does not show a clear source-flux dependence
in the present study.

\subsection{Comparison with the Previous FIR All-sky Map}

The FIS all-sky diffuse map provides 
an improved version of the FIR stacking study
that has been performed with the previous finest FIR all-sky 
map obtained with IRAS. 
As described in section \ref{resluts1},
the average FWHMs are less than $90\arcsec$ at the three bands 
and significantly smaller than those of the IRAS map ($\sim 6\arcmin$, \cite{beichman88}).
The higher angular resolution allows us to improve the sensitivity of faint source photometry 
by reducing confusion noise, which is one of the fundamental limiting factors
for FIR stacking studies. 
Therefore the FIS maps will significantly improve the quality of the studies of faint sources
performed with the IRAS data (e.g, \cite{kashiwagi12, bournaud12}). 

In addition to their sensitivities, the higher resolutions of the FIS maps enables 
us to study the spatial structures of the stacked sources,
such as debris disks around main-sequence stars \citep{thompson10}, 
and intergalactic dust associated with galaxies \citep{menard10} and galaxy clusters \citep{giard08}.
The angular scales of these objects are less than $10\arcmin$, which are 
difficult to resolve with the IRAS data.
With the {\it AKARI}/FIS all-sky maps, we can improve, 
and newly develop spatial studies 
of these faint sources by stacking analysis.   

\section{Summary}

Investigations of the PSFs and flux calibrations 
for faint sources in the {\it AKARI}/FIS all-sky 65, 90, and 140\,$\mu$m maps  
were performed, 
based on measurements of stacked calibration stars.
The PSFs 
are stable over a wide range of fluxes.
The deviation of the observed photometry for the stacked stars with different fluxes 
from the constant calibration factor
were estimated to be 9, 3, and 21\,\% at the 
65, 90, and 140\,$\mu$m bands, respectively.  
These results indicate that the performances are robust 
against flux even below the detection limits, 
and the FIS all-sky map is useful for the stacking analyses for faint sources.

These calibration results will be useful not only for the stacking analyses 
using the FIS all-sky data, but also for the absolute photometry 
for individual faint sources in the FIS diffuse map. 
The calibration scheme described in this paper 
will be also useful for the investigations of the data 
taken with the other FIR observations and future missions.

\bigskip
We thank the FIS instrument members 
for their valuable suggestion of calibration and 
data reduction.
This research is based on observations with {\it AKARI}, 
a JAXA project with the participation of ESA.
This research has made use of the SIMBAD database, 
operated at CDS, Strasbourg, France.
This work is partly supported by JSPS grants (2510319).



\newpage

\begin{table}
  \caption{Number of selected standard stars in each flux bin.}\label{taba2}
  \begin{center}
    \begin{tabular}{lccc}
      \hline
      flux range [Jy]& $ 60\, {\rm \mu m}$ &  $90\, {\rm \mu m}$ &  $140\, {\rm \mu m}$  \\
       \hline
      $> 3.98$          & 7     & 4    & 0 \\
      $1.58 - 3.98$   & 29   & 16   & 2 \\
      $0.63 - 1.58 $  & 41   & 43   & 11  \\
      $0.25 - 0.63$   & 88   & 39   & \multirow{2}{*}{70} \\
      $0.1 - 0.25$     & 175 & 182  &   \\	
      $< 0.1$            & 0     & 67    & 253  \\	
      \hline \\
    \end{tabular}
  \end{center}
\end{table}

\begin{table}
  \caption{FWHMs of stacked PSFs for the FIS all-sky map. \label{taba3}}
  \begin{center}
    \begin{tabular}{lcccc}
      \hline
      wavelength & $ 60\, {\rm \mu m}$ &  $90\, {\rm \mu m}$ &  $140\, {\rm \mu m}$   \\
       \hline
       FWHM & 53\arcsec & 73\arcsec & 86\arcsec \\
       In-scan FWHM   & 82\arcsec & 98\arcsec & 101\arcsec  \\
       Cross-scan FWHM  & 33\arcsec & 55\arcsec & 70\arcsec \\	
      \hline \\
    \end{tabular}
  \end{center}
\end{table}

\newpage

\newpage

\newpage

\newpage

\newpage

\newpage

\newpage


%
%



\begin{thebibliography}{}

\bibitem[Arimatsu et al.(2011)]{arimatsu11} Arimatsu, K., et al. 2011, \pasp, 123, 981

\bibitem[Beichman et al.(1988)]{beichman88} Beichman, C. A., Neugebauer, G., Habing, H. J., Clegg, P. E., \& Chester, T. J. 1988, Infrared Astronomical Satellite (IRAS) Catalogs and Atlases. Volume 1: Explanatory Supplement (Washington, DC: US GPO)

\bibitem[Bournaud et al.(2012)]{bournaud12} Bournaud, F. et al. 2012, \apj, 757, 19

\bibitem[Cohen et al.(1999)] {cohen99} Cohen, M., Walker, R. G., Carter, B., Hammersley, P., Kidger, M., \& Noguchi, K. 1999, \aj, 117, 1864

\bibitem[Giard et al.(2008)]{giard08} Giard, M., Montier, L., Pointecouteau, E., \& Simmat, E. 2008, \aap, 490, 547 

\bibitem[Jauzac et al.(2011)]{jauzac11} Jauzac, M. et al. 2011, \aap, 525, 52

\bibitem[Kaneda et al.(2002)]{kaneda02} 
Kaneda, H., Okamura, Y., Nakagawa, T., \& Shibai, H. 2002, Adv. Space Res., 30, 2105

\bibitem[Kashiwagi et al.(2012)]{kashiwagi12} Kashiwagi, T., Yahata, T., \& Suto, Y. et al.  2012, \pasj, 65, 12

\bibitem[Kawada et al.(2007)]{kawada07} Kawada, M. et al. 2007, \pasj, 59, 389

\bibitem[Matsuura et al.(2011)]{matsuura11} Matsuura, S., et al. 2011, \apj, 737, 2

\bibitem[M\'{e}nard et al.(2010)]{menard10} M\'{e}nard, B., Kilbinger, M., Scranton, R. 2010, MNRAS, 406, 1815

\bibitem[Murakami et al.(2007)]{murakami07} Murakami, H. et al. 2007, \pasj, 59, S369
 
\bibitem[Shirahata et al.(2009)]{shirahata09} Shirahata, M. et al. 2009, \pasj, 61, 737

\bibitem[Shirahata et al.(2004)]{shirahata04} Shirahata, M., et al. 2004, Proc. SPIE, 5487, 369

\bibitem[Thompson et al.(2010)]{thompson10} Thompson, M. A., et al. 2010. \aap, 518, L34

\bibitem[Yamamura et al.(2010)]{yamamura10} Yamamura, I., Makiuti, S., Ikeda, N., Fukuda, Y., Oyabu, S., Koga, T. \& White, G. J. 2010 \\
AKARI/FIS All-Sky Survey Bright Source Catalogue Version 1.0 Release Note \\
\url{http://irsa.ipac.caltech.edu/data/AKARI/documentation/AKARI-FIS\_BSC\_V1\_RN.pdf}

\end{thebibliography}
\end{document}